\documentclass{article}

\usepackage{amsfonts, amsmath, amssymb, amsbsy, graphicx}

\usepackage[sort,numbers]{natbib}

\usepackage{url} 

\begin{document}

\begin{center}
\textbf{\LARGE Variable stars identification in digitized photographic data}
\end{center}

\begin{center}
\textbf{K.~V.~Sokolovsky$^{1,2,3}$, 
D.~M.~Kolesnikova$^{4}$,
A.~M.~Zubareva$^{4,2}$, 
N.~N.~Samus$^{4,2}$,
S.~V.~Antipin$^{2,4}$}
\end{center}

\begin{center}
{\it
\noindent $^{1}$IAASARS, National Observatory of Athens, 15236 Penteli, Greece \\
$^2$Sternberg Astronomical Inst., Moscow State Uni., Universitetskii~pr. 13, 119992 Moscow, Russia \\
$^3$Astro Space Center, LPI RAS, Profsoyuznaya Str. 84/32, 117997 Moscow, Russia \\
$^4$Institute of Astronomy RAS, Pyatnitskaya Str. 48, 119017 Moscow, Russia \\ }
\end{center}

\begin{abstract}
We identify 339 known and 316 new variable stars of various types among 250000 
lightcurves obtained by digitizing 167 $30\times30$\,cm photographic plates of the
Moscow collection. We use these data to conduct a comprehensive test of 
18 statistical characteristics (variability indices) in search for the best
general-purpose variability detection statistic. We find that the highest
peak on the DFT periodogram, interquartile range, median absolute deviation, 
and Stetson's $L$ index are the most efficient in recovering variable objects 
from the set of photographic lightcurves used in our test.
\end{abstract}

\section{Introduction}

The simplest way to find a variable object is to compare its brightness on
two images of the sky taken at different times. However, this works well
only if the amplitude of brightness variations over that time is
large compared to measurement errors associated with the images.
If we have a lightcurve that includes measurements of an object's brightness
at multiple times, in principle, we may ``average out'' individual
measurement errors and recover a small-amplitude variability.
Two problems complicate this in practice: poor knowledge of measurement
errors (which is especially true for photographic data) and a~priori unknown 
pattern of object's variations. One may overcome the first problem by assuming
that objects that are close to each other in the sky and have similar brightness 
are measured with about the same accuracy on a given set of images.
To overcome the second problem one needs a variability indicator that
responds to a wide variety of brightness variation patterns.

In this work we compare 18 statistical characteristics (Table\,\ref{tab:indexsummary}) 
that quantify ``how variable'' an object is. 
The indices belong to three classes: 
{\it i)}\,scatter-based indices quantifying
the scatter of brightness measurements in a lightcurve; 
{\it ii)}\,correlation-based indices characterize the degree of correlation between
the consecutive brightness measurements;
{\it ii)}\,period-search methods look for periodic brightness variations.

The last column of Table\,\ref{tab:indexsummary} refers to the publications
in which one may find the definitions of these indices, so here we mention
only the more unconventional ones.
The interquartile range\footnote{\url{https://en.wikipedia.org/wiki/Interquartile_range}},
IQR \citep[e.g.][]{2014A&A...566A..43K} is a robust measure of scatter. 
It includes the inner 50\% of measurement values 
(i.e. excludes 25\% of the brightest and 25\% of the faintest flux measurements). 
Unlike the commonly used root mean square, the IQR is insensitive to outliers.
To use \cite{1965ApJS...11..216L} and \cite{1975ApSS..36..137D} period
search techniques as ``variability indices'' we compute the periodogram in
the 0.1--10\,d with steps in frequency corresponding to a phase shift of 0.01
between the first and the last points in a lightcurve. The value of the
highest peak on the periodogram is then used as a variability index.

\section{Comparison technique and results}

To test the performance of the variability indices we use
167 $30\times30$\,cm photographic plates ($10^\circ\times10^\circ$ field of view with a limiting magnitude
of $\sim17.5$\,pg) of the 104\,Her field. The plates are 
obtained with a 40\,cm $F=160$\,cm astrograph in 1976-1994, 
digitized with a flatbed scanner and split into 173 $52^\prime\times52^\prime$
partly overlapping subfields that were independently processed with
the {\scshape VaST}\footnote{\url{http://scan.sai.msu.ru/vast/}} software.
The lightcurves of 250000 stars were extracted and searched for variability
using the technique discussed by
\cite{2014aspl.conf...79S,2014ARep...58..319S,2010ARep...54.1000K,2008AcA....58..279K}.
The dataset includes 339 known and 316 new variable stars, among them 341~eclipsing
binaries, 165~RR~Lyrae stars and 139 red periodic, semi-periodic and
irregular variables. Having constructed the comprehensive list of true variable stars, 
we investigate how well these variables can be extracted from the dataset
using various variability indices.

To quantify the quality of candidate variables selection provided by each
variability index following 
\cite{2011ApJ...735...68K,2014MNRAS.439..703G,2016A&A...587A..18K}, we
compute the completeness $C$ and purity $P$:
\begin{equation}
C = \frac{\mathrm{Number \,\, of \,\, selected \,\,
variables}}{\mathrm{Total \,\, number \,\, of \,\, confirmed \,\,
variables}}
\label{eq:C}
\end{equation}
\begin{equation}
P = \frac{\mathrm{Number \,\, of \,\, selected \,\,
variables}}{\mathrm{Total \,\, number \,\, of \,\, selected \,\,
candidates}
\label{eq:P}}
\end{equation}
as well as the fidelity
$F_1$-score\footnote{The $C$ and $P$ parameters are often referred to as
``recall'' or ``sensitivity'' or ``true positive rate'' and ``precision'', 
respectively. See
\url{https://en.wikipedia.org/wiki/Precision_and_recall}}
which is the harmonic mean of the two parameters:
\begin{equation}
F_1 = 2 (C \times P) / (C + P).
\label{eq:F}
\end{equation}
$F_1 = 1$ for a perfect selection when all true
variables and no false candidates pass the selection criteria while $F_1=0$
if no true variables are selected. 

\begin{figure*}
 \centering
 \includegraphics[width=0.32\textwidth]{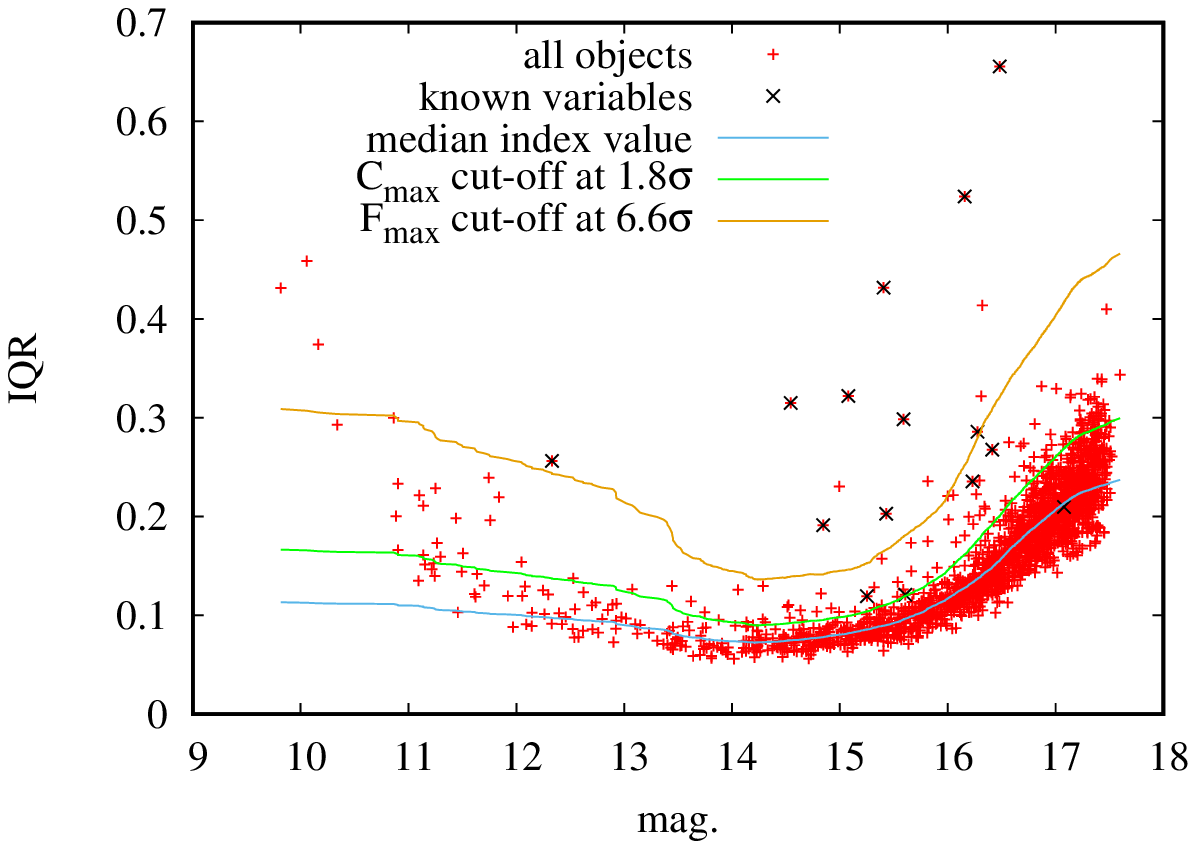}
 \includegraphics[width=0.32\textwidth]{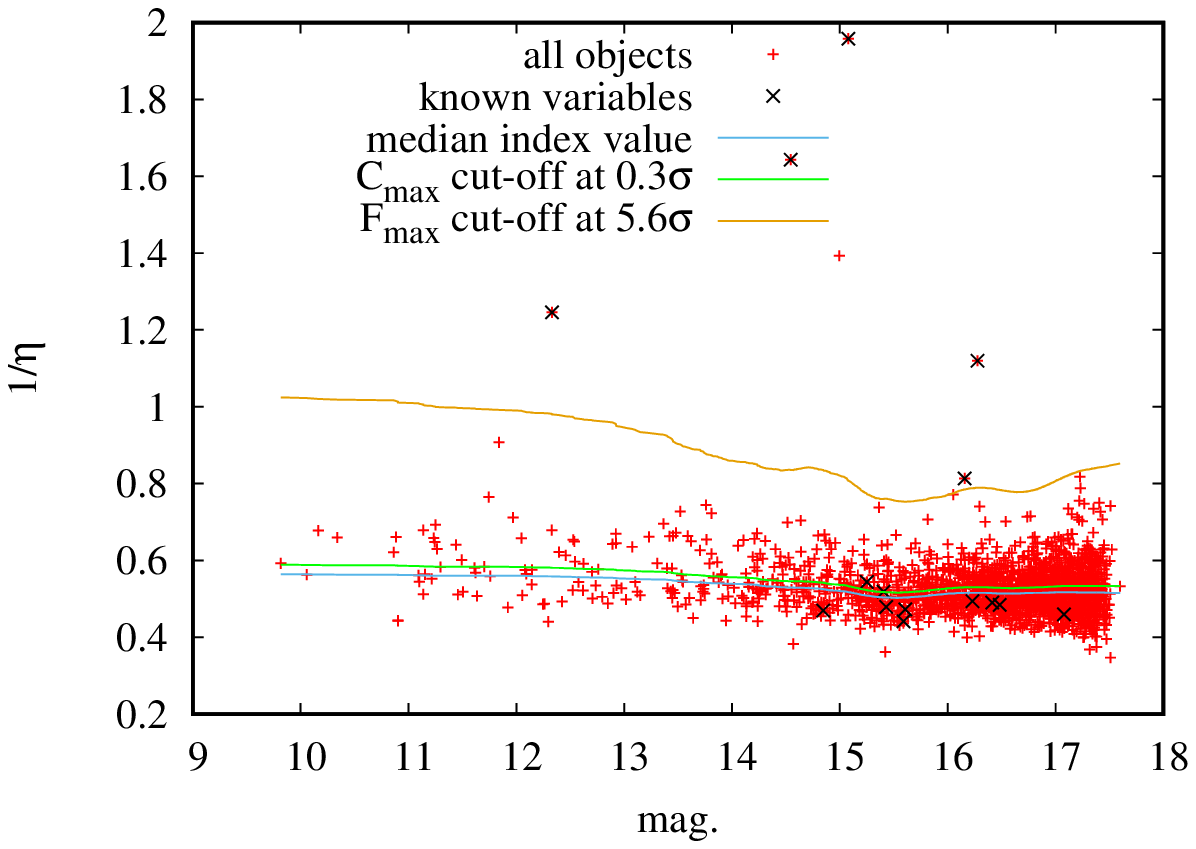}
 \includegraphics[width=0.32\textwidth]{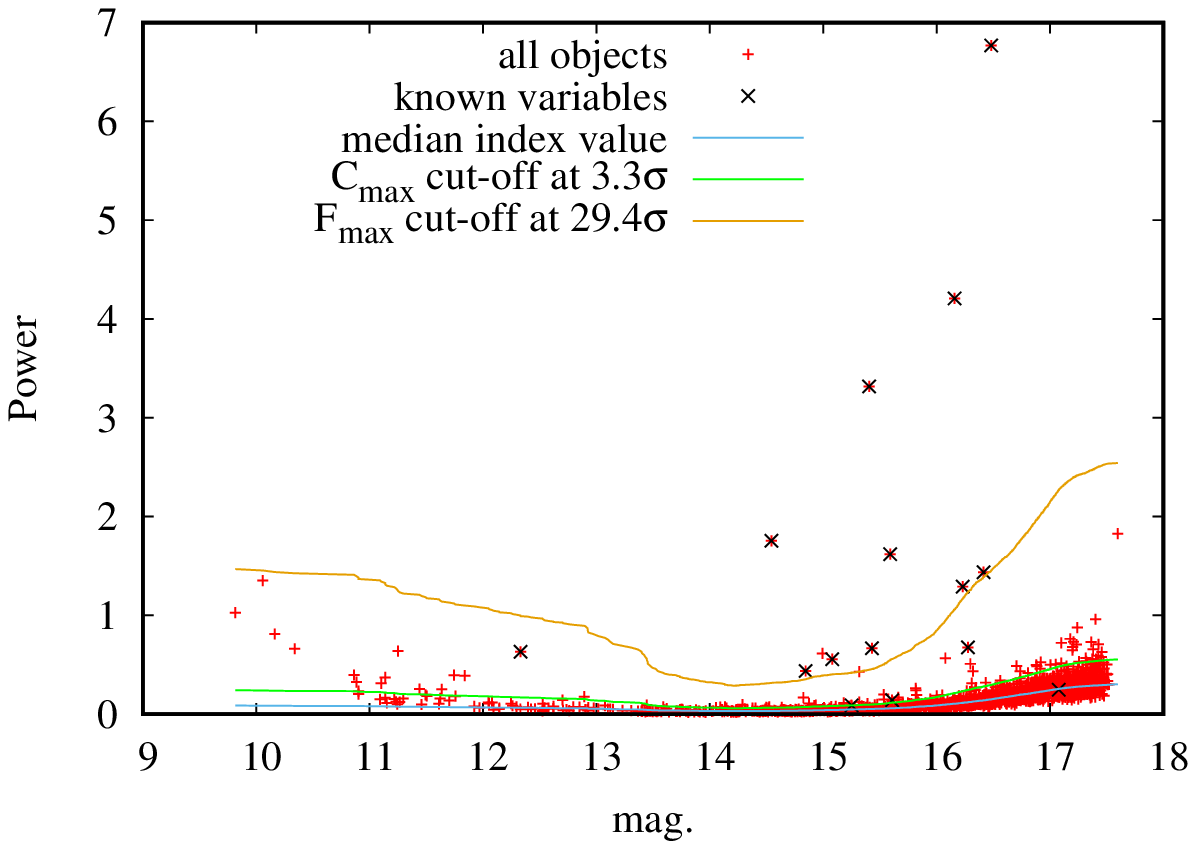} 
\caption{Variability indices IQR, $1/\eta$, and the highest DFT peak
plotted as a function of magnitude for one of the $52^\prime\times52^\prime$
subfields. Variable stars are marked with 'x'.
The curves represent the expected values of the indices for a given magnitude and
selection thresholds corresponding to the best trade-off between the completeness and purity of
the candidates list ($F_{\rm max}$) and the maximum completeness of the list
($C_{\rm max}$).}
 \label{fig:idx}
\end{figure*}

\begin{figure*}
 \centering
 \includegraphics[width=0.32\textwidth]{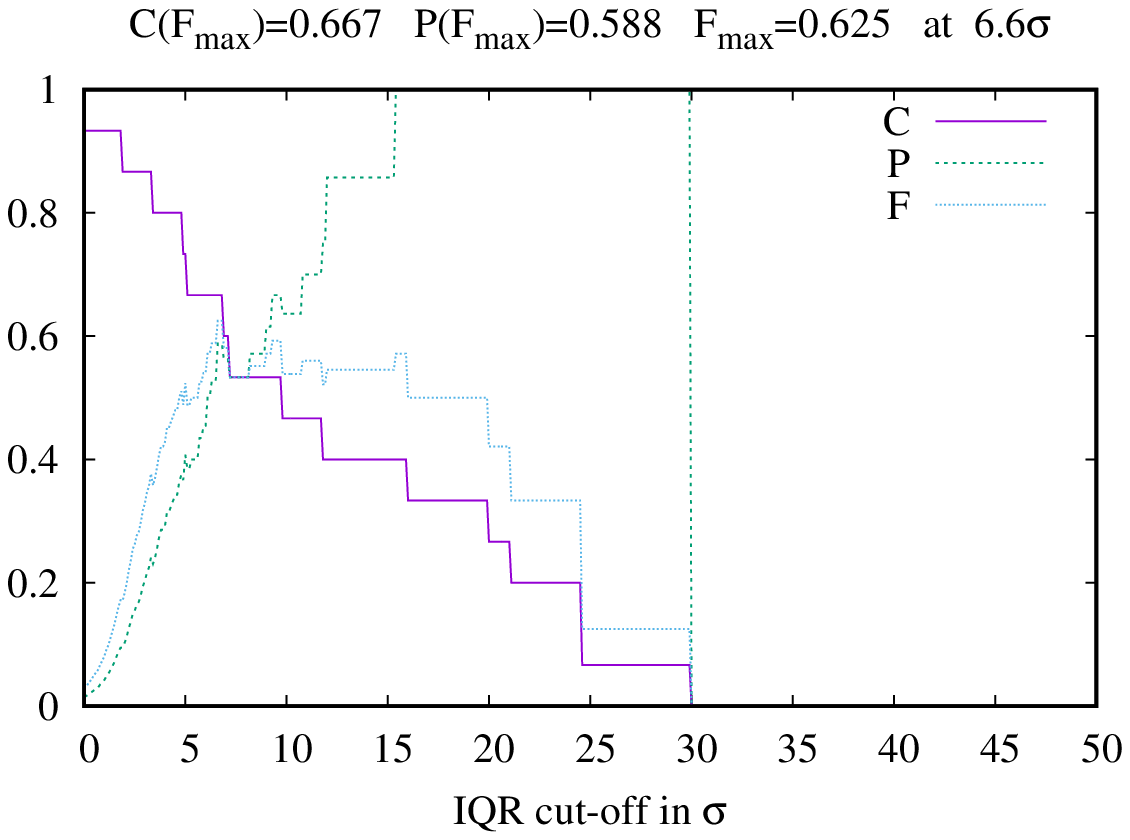}
 \includegraphics[width=0.32\textwidth]{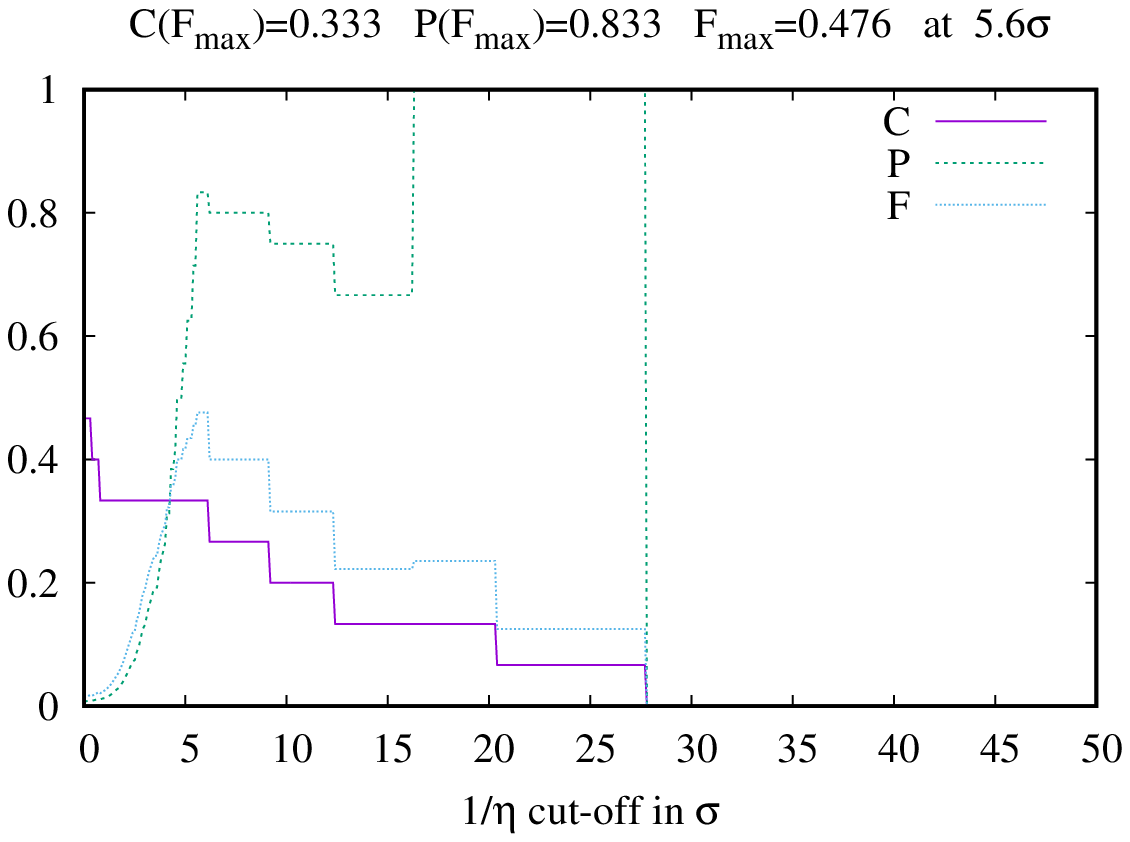}
 \includegraphics[width=0.32\textwidth]{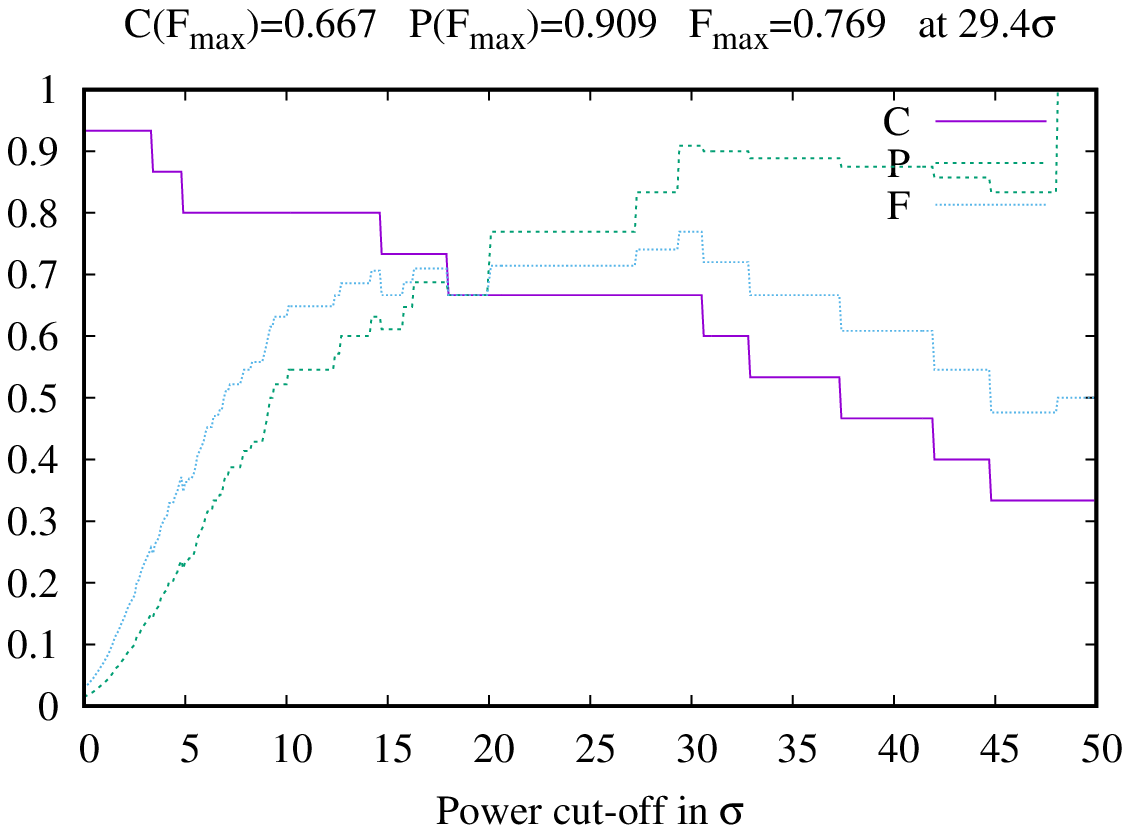} 
\caption{Variable star selection completeness ($C$, Eq.\,\ref{eq:C}),
purity ($P$, Eq.\,\ref{eq:P}), and $F_1$-score (Eq.\,\ref{eq:F}) as a function of selection threshold for the
variability indices 
IQR, $1/\eta$, and the highest DFT peak.
The plots are for the dataset presented at Fig.\,\ref{fig:idx}.}
 \label{fig:CPF}
\end{figure*}

For each variability index we estimate its expected value and 
its scatter, $\sigma$, as a function of magnitude (Fig.\,\ref{fig:idx}).
Candidate variables are then selected as objects having their variability 
index value $>n\sigma$ above the expected value of this index for the
object's magnitude. The selection is repeated for $n$ in the range 0--50.
The resulting $C$, $P$, and $F_1$ values as a function of $n$ are presented
in Fig.\,\ref{fig:CPF}.
The selection resulting in the highest $F_1$-score is used to compare the indices. 
This way the optimal cut-off value $n\sigma$ is used for each index. 
The distribution of the expected index values for a given magnitude is non-Gaussian, 
therefore a simple choice like a $3\sigma$ cut-off might not be the optimal one 
for some indices.

The results of variability indices comparison are presented in
Table\,\ref{tab:indexsummary}. The table presents the information 
taken into account by each index (in addition to the measured magnitudes 
themselves) that may include estimated photometric errorbars, order of points in a
lightcurve and exact times of observations. It presents the maximum
$F_1$-score reached by a selection using each index.
We consider the index with the highest value of $F_{1~\rm max}$ as the most
efficient in selecting true variable stars. Since $F_1$ characterizes only
the selected candidates, but does not take into account the rejected,
presumably non-variable, objects, 
Table\,\ref{tab:indexsummary} also lists a fraction of objects that do not pass the selection 
(at the cut-off value corresponding to $F_{1~\rm max}$), $R_{F_{1~\rm max}}$, as
an auxiliary measure of variability index performance. Finally,
Table\,\ref{tab:indexsummary} reports the maximum completeness, 
$C_{\rm max}$, reached by each index at a selection cut-off of $n\sigma$
where $n \ge 0$. The values of $C_{\rm max}<1$ indicate that the index
cannot recover some variable stars, even at a low selection threshold
(corresponding to a large number of false candidates).
All $F_{1~\rm max}$, $R_{F_{1~\rm max}}$, and $C_{\rm max}$ values presented in
Table\,\ref{tab:indexsummary} are the median values computed over the
173 subfields.

\section{Conclusions}

Table\,\ref{tab:indexsummary} indicates that the highest peak on the DFT
periodogram, ${\rm IQR}$, ${\rm MAD}$, and Stetson's $L$ index are the most
efficient in recovering variable objects from the set of photographic 
lightcurves used for the test. These indices can be recommended for the
future searches of variable objects using photographic lightcurves.
Some correlation-based indices (like the $I$ index) 
are only able to recover objects varying on 
timescales longer than the typical lightcurve sampling time and, therefore,
are not good general-purpose variability indicators for (typically) sparsely 
sampled photographic lightcurves.
Constant stars with corrupted measurements (e.g. due to blending with a
nearby star) may pass the selection threshold even for the best identified
variability indices. The need to reject such badly measured stars through a
visual inspection of lightcurves and images so far prevents a full automation 
of variability searches.



\begin{table}
\begin{center}
   \caption{Variability indices}
   \label{tab:indexsummary}
   \begin{tabular}{c c@{~~}c@{~~}c c@{~~}c@{~~}c c}
   \hline\hline
Index                     & Errors & Order & Time  & $F_{1~\rm max}$ & $R_{F_{1~\rm max}}$ & $C_{\rm max}$ & Ref. \\
   \hline
\multicolumn{8}{c}{Scatter-based indices} \\
$\chi_{\rm red}^2$        & $\checkmark$ &              &              & 0.111 & 0.979 & 1.000 & \cite{2010AJ....139.1269D} \\
$\sigma$                  &              &              &              & 0.182 & 0.987 & 1.000 & \cite{2008AcA....58..279K} \\
${\rm MAD}$               &              &              &              & 0.400 & 0.995 & 1.000 & \cite{2016PASP..128c5001Z} \\
${\rm IQR}$               &              &              &              & 0.400 & 0.995 & 1.000 & this work \\
${\rm RoMS}$              & $\checkmark$ &              &              & 0.333 & 0.994 & 1.000 & \cite{2007AJ....134.2067R} \\
$\sigma_{\rm NXS}^2$      & $\checkmark$ &              &              & 0.200 & 0.990 & 1.000 & \cite{1997ApJ...476...70N} \\
$v$                       & $\checkmark$ &              &              & 0.039 & 0.932 & 1.000 & \cite{2009AN....330..199S} \\
\multicolumn{8}{c}{Correlation-based indices} \\
$l_1$                     &              & $\checkmark$ &              & 0.250 & 0.997 & 0.667 & \cite{2011ASPC..442..447K} \\
$I$                       & $\checkmark$ & $\checkmark$ & $\checkmark$ & 0.154 & 0.989 & 0.667 & \cite{1993AJ....105.1813W} \\
$J$                       & $\checkmark$ & $\checkmark$ & $\checkmark$ & 0.250 & 0.994 & 1.000 & \cite{1996PASP..108..851S} \\
$J({\rm time})$           & $\checkmark$ & $\checkmark$ & $\checkmark$ & 0.250 & 0.995 & 0.750 & \cite{2012AJ....143..140F} \\
$L$                       & $\checkmark$ & $\checkmark$ & $\checkmark$ & 0.400 & 0.996 & 1.000 & \cite{1996PASP..108..851S} \\
$E_x$                     & $\checkmark$ & $\checkmark$ & $\checkmark$ & 0.222 & 0.993 & 1.000 & \cite{2014ApJS..211....3P} \\
$1/\eta$                  &              & $\checkmark$ &              & 0.250 & 0.998 & 0.667 & \cite{2009MNRAS.400.1897S} \\
$\mathcal{E}_\mathcal{A}$ &              & $\checkmark$ & $\checkmark$ & 0.014 & 0.860 & 0.600 & \cite{2014AnA...568A..78M} \\
$S_B$                     & $\checkmark$ & $\checkmark$ &              & 0.143 & 0.987 & 1.000 & \cite{2013AnA...556A..20F} \\
\multicolumn{8}{c}{Period search} \\
${\rm L-K}$               &              & $\checkmark$ & $\checkmark$ & 0.087 & 0.981 & 1.000 & \cite{1965ApJS...11..216L} \\
${\rm DFT}$               &              & $\checkmark$ & $\checkmark$ & 0.500 & 0.995 & 1.000 & \cite{1975ApSS..36..137D} \\
   \hline
   \end{tabular}
   \renewcommand{\arraystretch}{1.0}
\end{center}
\end{table}

{\small {\bf Acknowledgements.} 
KVS is supported by the European Space Agency (ESA) under
the ``Hubble Catalog of Variables'' program, contract No.\,4000112940.
This work is supported by the grant from the Program ``Transition and
explosive processes in the Universe'' of the Presidium      
of Russian Academy of Sciences and the RFBR grant 13-02-00664.}

\bibliographystyle{Science}
\bibliography{sokolovsky_vssearch16}

\end{document}